 \documentclass[aps,pra,superscriptaddress,amsmath,amssymb,preprintnumbers,twocolumn,floatfix,showpacs,showkeys,10pt]{revtex4-1}
 \usepackage{amssymb} \usepackage{epsfig}
 \begin{document}
  \title{Quantum coherence of two-qubit over quantum channels with memory  }

\author{You-neng Guo}
\email{guoxuyan2007@163.com}
\affiliation{ Department of Electronic and Communication Engineering, Changsha University, Changsha, Hunan
410022, People's Republic of China}
\affiliation{Hunan Province Key Laboratory of Applied Environmental Photocatalysis, Changsha University, Changsha, Hunan
410022, People's Republic of China}
\author{Ke Zeng}
\email{zk92@126.com}
\affiliation{ Department of Electronic and Communication Engineering, Changsha University, Changsha, Hunan
410022, People's Republic of China}
\affiliation{Hunan Province Key Laboratory of Applied Environmental Photocatalysis, Changsha University, Changsha, Hunan
410022, People's Republic of China}
\author{Qing-long Tian}
\affiliation{ Department of Mathematics and Computing Science, Changsha University, Changsha, Hunan
410022, People's Republic of China}
\author{Zheng-da Li}
\email{correspond author: 0736a@163.com}
\affiliation{ Department of Electronic and Communication Engineering, Changsha University, Changsha, Hunan
410022, People's Republic of China}
\affiliation{Hunan Province Key Laboratory of Applied Environmental Photocatalysis, Changsha University, Changsha, Hunan
410022, People's Republic of China}

\begin{abstract}
Using the axiomatic definition of the coherence measure, such as the $l_{1}$ norm and the relative entropy, we study the phenomena of two-qubit system quantum coherence through quantum channels where successive uses of the channels are memory. Different types of noisy channels with memory, such as amplitude damping, phase-damping, and depolarizing channels effect on quantum coherence have been discussed in detail. The results show that, quantum channels with memory can efficiently protect coherence from noisy channels. Particularly, as channels with perfect memory, quantum coherence is unaffected by the phase damping as well as depolarizing channels. Besides, we  also investigate the cohering and decohering power of quantum channels with memory.

 \end{abstract}

 \maketitle
\section{Introduction}
Quantum coherence not only is the key resource in quantum information theory, but also becomes a basic feature of quantum physics in quantum physics~\cite{1,2,3,4,5}. Usually, quantum coherence~\cite{1}, which is identified by the absolute values of off-diagonal terms in the quantum states, is ascribed to the superposition principle in quantum physics. As the resource theory of coherence proposed, the study on quantum coherence of characterization and quantification has attracted much attention~\cite{6,6a,7,7a,8,8a,9,9a,10,10a}. Compared with quantum correlations, however quantum coherence has its dual characters. On one hand, like quantum correlations, coherence is very fragile and prone to environmental effect due to any realistic systems inevitably interacting with their external environments. This means that, quantum coherence is usually very difficult to be created, maintained, and manipulated in quantum systems.
Therefore, it is very important and significant to create and preserve the quantum coherence in the field of quantum information processing and quantum computation.

On the other hand, different from quantum correlations, quantifying of quantum coherence depends on a reference basis chosen.  For example, a qubit state like $|+\rangle =\frac{\sqrt{2}}{2}(|0\rangle + |1\rangle)$ possesses maximum coherence in the basis $\{|0\rangle, |1\rangle\}$, however it is not true in the basis $\{|+\rangle, |-\rangle\}$ in which it has zero coherence. In order to exploit quantum coherence, one needs to quantify coherence for a given state. Fortunately, Baumgratz et al.~\cite{11} first introduced a rigorous framework for the resource theory of coherence including the definition of incoherent states, incoherent operations and arbitrary valid coherence measures. They suggested any proper measure of the coherence must satisfy several criteria. Along this line, fruitful research has been made, some of which was mainly focused on studying the properties of specific coherence measures, for example, skew-information of coherence has been introduced in Ref.~\cite{12}. The coherence measures based on entanglement~\cite{13,14}, operation~\cite{15,16}, and convex-roof construction~\cite{17,18} were subsequently proposed, and another of which was mainly devoted to investigating quantum coherence dynamics and its protection~\cite{17a,17b,17c,18a,18b}. In this paper, we adopt two basic concrete measures to study the phenomena of two-qubit system quantum coherence suffered from quantum correlated channels, one based on the $l_{1}$-norm of off-diagonal elements and the other based on the relative entropy of coherence. The $l_{1}$-norm of coherence $C_{l_1}$ is defined as~\cite{11}
\begin{equation}\label{Eq1}
C_{l_1}(\rho)=\min\limits_{\delta \in \mathcal{I}}D(\rho,\delta)=\sum\limits_{i\neq j}|\rho_{ij}|,
\end{equation}
where $D(\rho,\delta)=\|\rho-\delta\|_{1}$ denotes the minimal distance of $\rho$ to a set of incoherent states $\mathcal{I}$ $({\delta \in \mathcal{I}})$. And $\rho_{ij}$ is the off-diagonal element of a quantum state $\rho$. The relative entropy of coherence $C_{R}$ is defined as~\cite{11}
\begin{equation}\label{Eq2}
C_{R}(\rho)=\min\limits_{\delta \in \mathcal{I}}S(\rho\|\delta)=S(\rho_{diag})-S(\rho).
\end{equation}
where $\rho_{diag}$ is obtained by removing all the off-diagonal elements in the density operator $\rho$, and $S(\rho)$ is the von Neumann entropy. It is quite easy to check that both of measures satisfy the conditions of coherence measures which are proposed by Baumgratz et al.~\cite{11}. In general for a $d$-dimensional
system they satisfy $0 \leq C_{l_1}(\rho)\leq d-1$ and $0 \leq C_{R}(\rho) \leq \log_{2}d$, respectively. Note that, even though there are closed form solutions that make them easy to evaluate analytical expressions, they still exist slightly different quantitatively.

As mention above that quantum coherence is a useful physical resource but complex systems are inevitably subject to noise which leads to destroy the coherence.  A challenge in exploiting the resource is how to protect quantum coherence from the decoherence caused by noise, and under which dynamical conditions the quantum coherence is unaffected by noise.
Most recently, Bromley et al. [19] found that under some suitable dynamic conditions, the quantum coherence in an open quantum system is totally
unaffected by noise during the entire evolution. Inspired by their results, Yang et al.[20] studied universal conditions where some coherence measures are amplified and frozen under
identical bit-flip channels by using quantum weak measurements or quantum measurement reversals. Here, we use the axiomatic definition of two coherence measure, namely, the $l_{1}$ norm and the relative entropy, to analyze under which dynamical conditions the quantum coherence of two-qubit system is unaffected by the correlated channels. The results show that, the quantum coherence of two-qubit system suffered from correlated noise can be well protected. The stronger the memory coefficient of channel is, the much more slowly the quantum coherence decay.  Particularly, as channels with perfect memory, quantum coherence is unaffected by the phase damping as well as depolarizing channels. Finally, we discuss the cohering and de-cohering powers of correlated channels.

The layout is as follows: In Sec. \textrm{II}, we first review noisy channels with memory and initial states, and then using two basic concrete measures to investigate the quantum coherence dynamics suffered from correlated noise. In Sec. \textrm{III}, we study the cohering power and decohering power of correlated channels and derive their exact analytical expressions on the fixed basis. Finally, we give the conclusion in Sec. \textrm{IV}.

\section{Initial states and noise model}  
In what follows, we begin with a brief description of quantum-memory
channels. Given an input state $\rho$, a quantum channel $\varepsilon$ is defined as a
completely positive, trace-preserving map from input-state
density matrices to output-state density matrices,
\begin{equation}
\varepsilon(\rho)=\sum_{i}E_{i}\rho E_{i}^{\dagger}
\end{equation}
where $E_{i}=\sqrt{P_{i_{1}...i_{N}}}A_{i}$ are the Kraus operators of the channel which satisfy the completeness relationship, and $\sum_{i}P_{i_{1}...i_{N}}=1$. $P_{i_{1}...i_{N}}=P_{i_{1}}P_{i_{2}}...P_{i_{N}}$ expresses the joint probability that a
random sequence of operations is applied to the sequence of
$N$ qubits transmitted through the memoryless channels. However for a channel with memory, $P_{i_{1}...i_{N}}=P_{i_{1}}P_{i_{2}|i_{1}}...P_{i_{N}|i_{N-1}}$, here $P_{i_{N}|i_{N-1}}$ is the conditional probability.

Following, we will focus on the noisy channels with time-correlated Markov noise for two consecutive uses. Based on the Kraus operator approach, two consecutive uses correlated channel maps input states $\rho$ onto~\cite{20a}
\begin{equation}\label{Eq3}
\varepsilon(\rho)=(1-\mu)\sum_{i,j}E_{i,j}\rho E_{i,j}^{\dagger}+\mu\sum_{k}E_{k,k}\rho E_{k,k}^{\dagger}
\end{equation}
the Kraus operators for two consecutive uses of a channel with memory are
\begin{equation}
E_{i,j}=\sqrt{P_{i}[(1-\mu)P_{j}+\mu\delta_{i,j}]}A_{i}\otimes A_{j}
\end{equation}
where $0\leq \mu \leq 1$ is the memory coefficient of channel. The limiting cases $\mu=0$ and $\mu=1$ correspond to memoryless and channels with perfect memory. It is evident from the above expression that the same operation is
applied to both qubits with probability $\mu$ while with probability $1-\mu$ both operations are uncorrelated. For simplicity,
the input states are taken the form as following
\begin{equation}\label{Eq4}
\rho=\frac{1}{4}(I\otimes I+\Sigma_{i=1}^{3}c_{i}\sigma_{i}\otimes \sigma_{i})
\end{equation}
where $0\leq|c_{i}|\leq 1$, $I$ is the identity operator on the subsystem, and $\sigma_{i}$ $(i=1,2,3)$ are the Pauli operators.

\begin{figure}[htpb]
  \begin{center}
   \includegraphics[width=7.5cm]{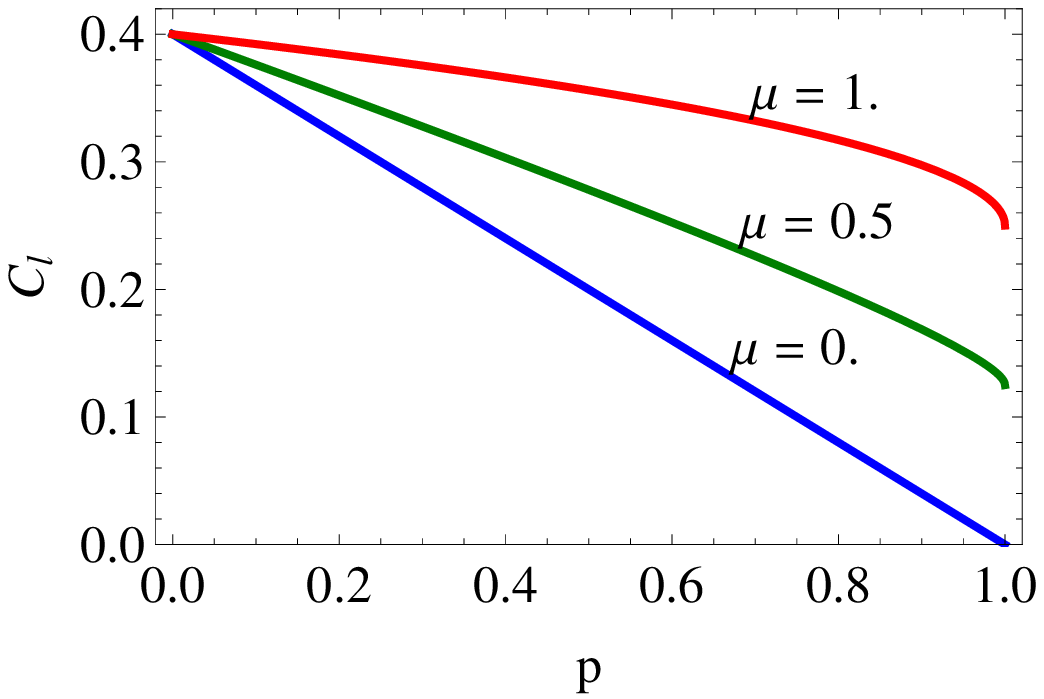}
   \includegraphics[width=7.5cm]{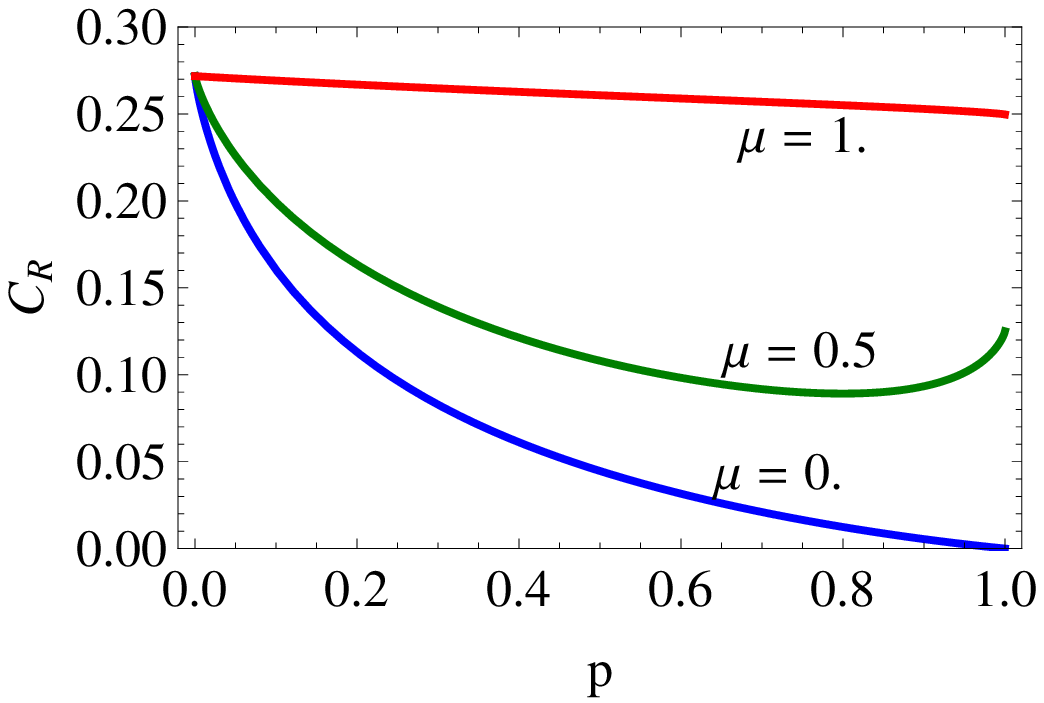}
 \caption{\label{Fig2}(Color online) The evolution of $l_{1}$-norm of coherence $C_{l_1}$  and relative
entropy of coherence $C_{R}$ for a two-qubit system passes through the amplitude damping channels with memory. Other parameters: $c_{1}=0.1,c_{2}=0.4,c_{3}=0.5$.}
\end{center}
\end{figure}
\begin{figure}[htpb]
  \begin{center}
   \includegraphics[width=7.5cm]{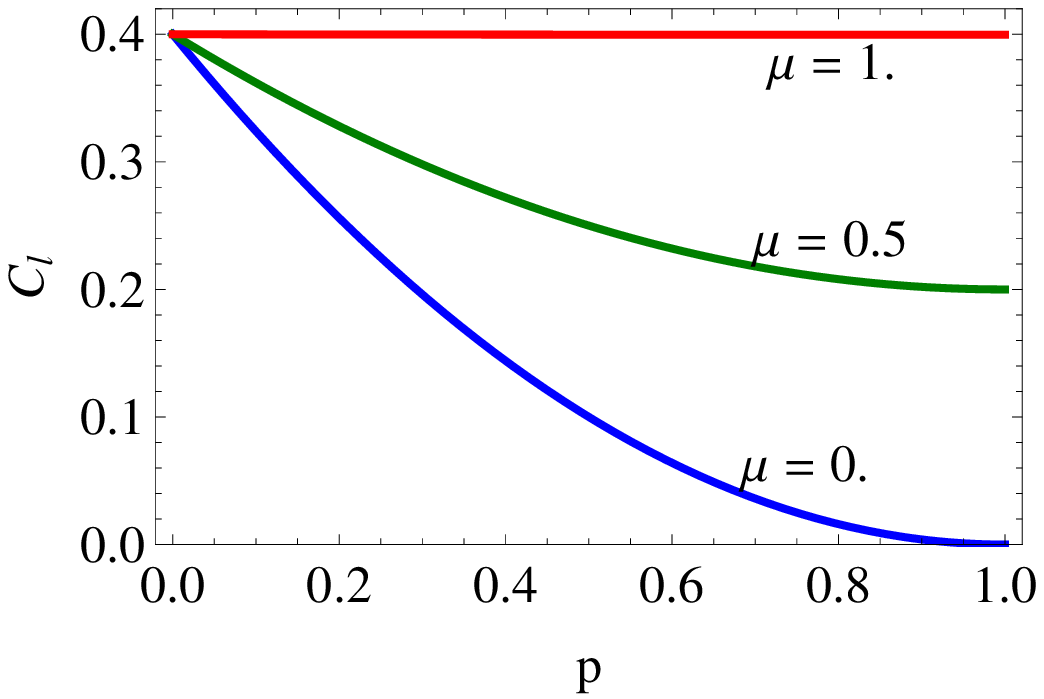}
   \includegraphics[width=7.5cm]{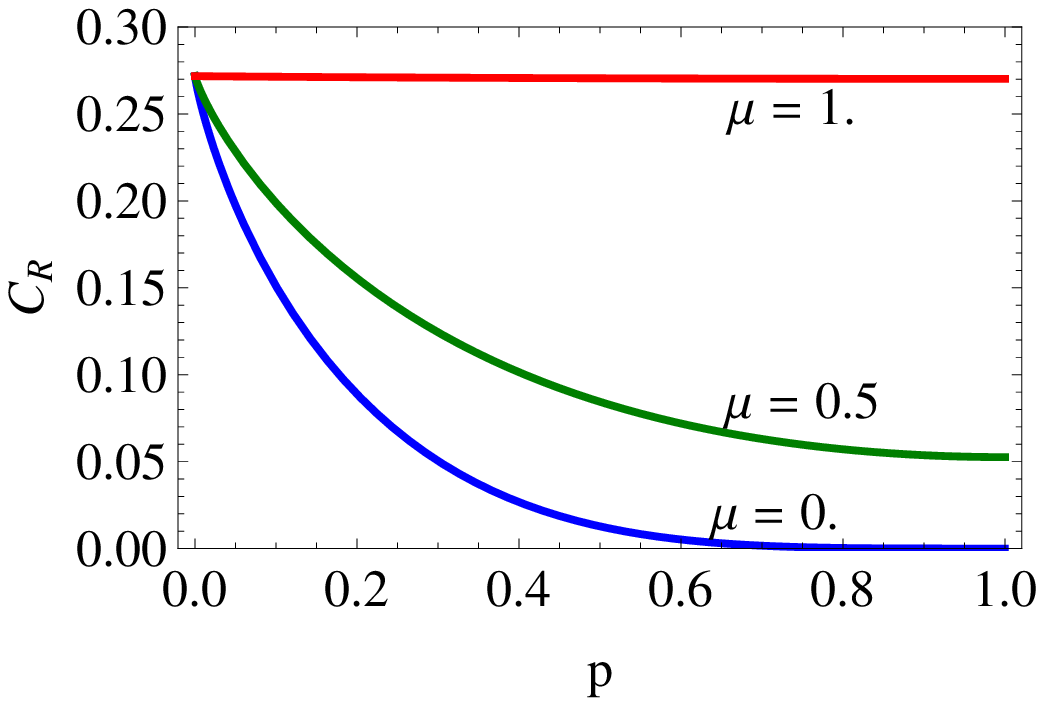}
 \caption{\label{Fig3}(Color online) The evolution of $l_{1}$-norm of coherence $C_{l_1}$  and relative
entropy of coherence $C_{R}$ for a two-qubit system passes through the phase damping channels with memory. Other parameters: $c_{1}=0.1,c_{2}=0.4,c_{3}=0.5$.}
\end{center}
\end{figure}
\begin{figure}[htpb]
  \begin{center}
   \includegraphics[width=7.5cm]{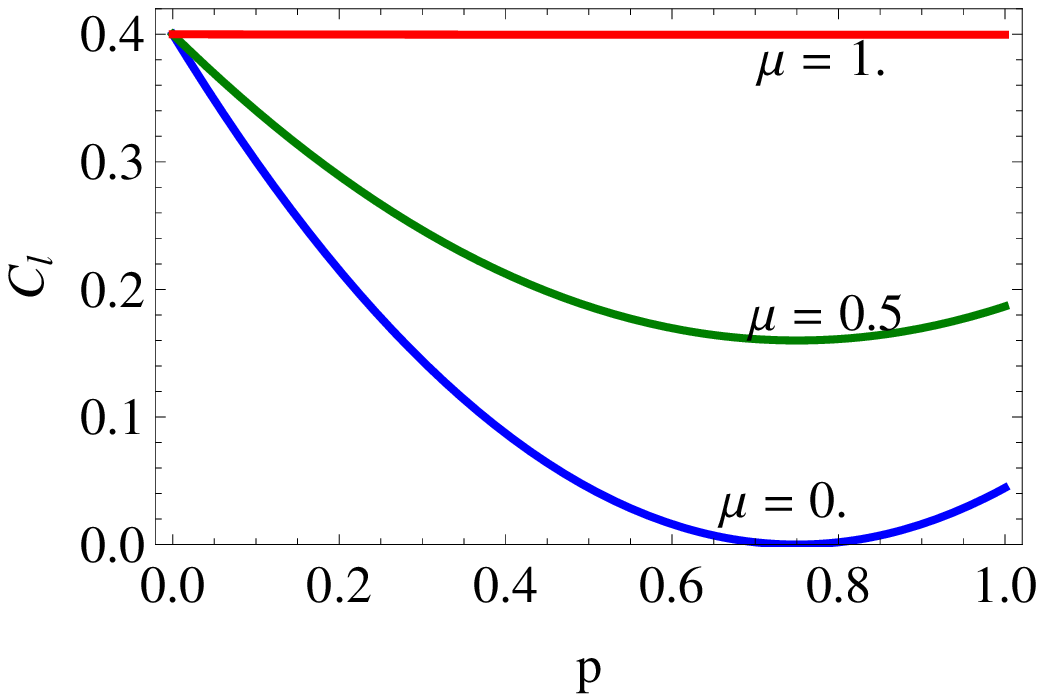}
   \includegraphics[width=7.5cm]{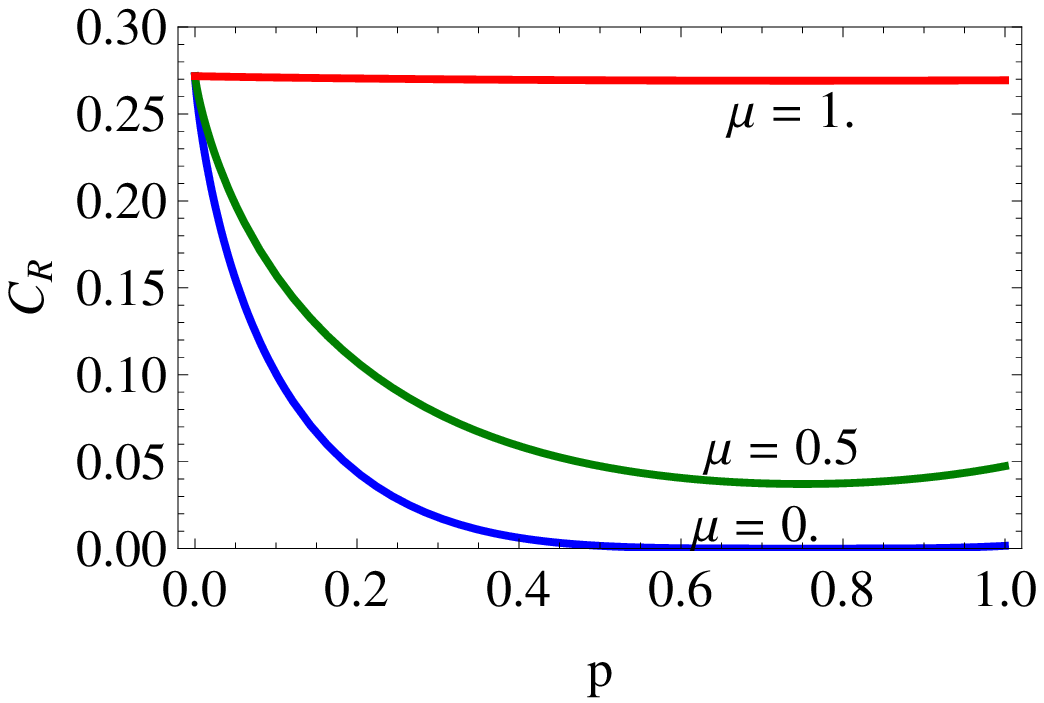}
 \caption{\label{Fig4}(Color online) The evolution of $l_{1}$-norm of coherence $C_{l_1}$  and relative
entropy of coherence $C_{R}$ for a two-qubit system passes through the depolarizing damping channels with memory. Other parameters: $c_{1}=0.1,c_{2}=0.4,c_{3}=0.5$.}
\end{center}
\end{figure}

\subsection{Amplitude damping channel with memory}
Amplitude damping channel which is used to characterize spontaneous emission describes the energy dissipation from a quantum system. The Kraus operators for a single qubit are given by~\cite{20b}
\begin{equation}A_{0}= \left(
\begin{array}{ c c c c l r }
\sqrt{1-p} & 0  \\
0 & 1  \\
\end{array}
\right)
\end{equation}
\begin{equation}A_{1}= \left(
\begin{array}{ c c c c l r }
0 & 0  \\
\sqrt{p} & 0  \\
\end{array}
\right)
\end{equation}
where $p=1-\exp(-\gamma t)$. Consider two consecutive uses quantum amplitude damping channel, the Kraus operators of channel with memoryless are
\begin{equation}\label{Eq5}
E_{i,j}=A_{i}\otimes A_{j},(i,j=0,1)
\end{equation}
and the Kraus operators of channel with memory are
\begin{equation}E_{00}= \left(
\begin{array}{ c c c c l r }
\sqrt{1-p} & 0 & 0 & 0  \\
0 & 1 & 0 & 0  \\
0& 0 & 1& 0 \\
0& 0 & 0 & 1\\
\end{array}
\right)
\end{equation}
\begin{equation}E_{11}= \left(
\begin{array}{ c c c c l r }
0 & 0 & 0 & 0  \\
0 & 0 & 0 & 0 \\
0 & 0 & 0 & 0 \\
\sqrt{p} & 0 & 0 & 0  \\
\end{array}
\right)
\end{equation}

According to Eq.~\eqref{Eq3}, output states of the amplitude-damping channel with memory for two consecutive uses,
\begin{equation}\label{Eq6}
\varepsilon(\rho)=\frac{1}{4}(I\otimes I+a_{3}^{'}\sigma_{3}\otimes I+b_{3}^{'}\sigma_{3}\otimes I+\Sigma_{i=1}^{3}c_{i}^{'}\sigma_{i}\otimes \sigma_{i})
\end{equation}
where

$c_{1}^{'}=\frac{1}{2}\{c_{2}\mu_{0}+ c_{1}[2-2p(1-\mu)-\mu_{0}]\}$,

$c_{2}^{'}=\frac{1}{2}\{c_{1}\mu_{0}+ c_{2}[2-2p(1-\mu)-\mu_{0}]\}$,

$c_{3}^{'}=c_{3}[1-(2-p)p(1-\mu)]+p^2(1-\mu)$,

$a_{3}^{'}=\frac{1}{2} [-2-\mu (1-c_{3})]p$,

$b_{3}^{'}=\frac{1}{2} [-2-\mu (1-c_{3})]p$,

$\mu_{0}=(1-\sqrt{1-p})\mu$.

Substituting Eq.~\eqref{Eq6} into Eq.~\eqref{Eq1} and Eq.~\eqref{Eq2} respectively,  the $l_{1}$ norm
and relative entropy of coherence for two qubits going through the amplitude-damping channel with memory are obtained
\begin{eqnarray}\label{Eq7}
C_{l_1}(\varepsilon(\rho))&=&\frac{1}{2}\{|c_{1}-c_{2}|[1-p(1-\mu)-(1-\sqrt{1-p})\mu]\nonumber\\
&+&|c_{1}+c_{2}|[1-p(1-\mu)]\},
\end{eqnarray}
and
\begin{eqnarray}\label{Eq8}
&&C_{R}(\varepsilon(\rho)) \nonumber\\
&=&\frac{1}{4}\{(1+c_{3})(p-1)[1-p(1-\mu)]\nonumber\\
&\times&\log_{2}\frac{1}{4}(1+c_{3})(1-p)[1-p(1-\mu)]\}\nonumber\\
&+&2[(1-p)(1-\mu)(1-c_{3}+c_{3}p+p)+(1-c_{3})\mu]\nonumber\\
&\times&\log_{2}\frac{1}{4}[(1-p)(1-\mu)(1-c_{3}+c_{3}p+p)+(1-c_{3})\mu]\nonumber\\
&+&[(c_{3}(1-p)^2+(1+p)^2)(1-\mu)+(1+c_{3})(1+p)\mu]\nonumber\\
&\times&\log_{2}\frac{1}{4}[(c_{3}(1-p)^2+(1+p)^2)(1-\mu)+(1+c_{3})(1+p)\mu]\nonumber\\
&+&\lambda_{1}\log_{2}\lambda_{1}+\lambda_{2}\log_{2}\lambda_{2}+\lambda_{3}\log_{2}\lambda_{3}+\lambda_{4}\log_{2}\lambda_{4},
\end{eqnarray}
where $\lambda_{i}$ are eigenvalues of the density matrix $\varepsilon(\rho)$ given by Eq.~\eqref{Eq6}.

To demonstrate quantum coherence properties of two qubits affected by the amplitude damping channels with memory, we firstly consider the case where the input states are given by Eq.~\eqref{Eq4}, (e.g.$c_{1}=0.1,c_{2}=0.4,c_{3}=0.5$) when two qubits system pass through the quantum channels with memory. We plot the evolutions of $l_{1}$ norm and relative entropy as the function of $p$ for different parameters $\mu$ in Fig. 1. It can be seen quantum coherence dynamics depend on the memory coefficient of channel $\mu$. The stronger the memory coefficient of channel $ \mu$ is, the much more slowly the quantum coherence decay. This implies the amplitude damping channels with memory can protect quantum coherence effectively. Particularly, there exists long-live quantum coherence in the limited $p\rightarrow 1$, namely time $t\rightarrow \infty$. One can easily determine from Eqs.~(\ref{Eq7}) and ~(\ref{Eq8}) in the limited $p\rightarrow 1$ which reduce to
\begin{eqnarray}\label{Eq7a}
C_{l_1}(\varepsilon(\rho))&=&\frac{1}{2}|c_{1}+c_{2}|\mu
\end{eqnarray}
and
\begin{eqnarray}\label{Eq7b}
C_{R}(\varepsilon(\rho))&=&\frac{1}{4}[\mu(1+c_{1}+c_{2}-c_{3})\log_{2}\mu(1+c_{1}+c_{2}-c_{3})\nonumber\\
&+&\mu(1-c_{1}-c_{2}-c_{3})\log_{2}\mu(1-c_{1}-c_{2}-c_{3})\nonumber\\
&-&2\mu(1-c_{3})\log_{2}\mu(1-c_{3})],
\end{eqnarray}
respectively. We can find that quantum coherence is only dependent of the initial input states parameters $c_{i}$ and the memory coefficient of channel $ \mu$. What one should expect is whether the quantum coherence of any input states is unaffected by the amplitude damping channels.
We argue that this case usually is not ture. According to Eqs.~\eqref{Eq7} and ~\eqref{Eq8}, we can obtain if only if the results $c_{1}=c_{2}=0$, the quantum coherence dynamics of a two qubits system over the amplitude damping channels with memory is unaffected by noisy channel. This results manifest the amplitude damping channels with memory must destroy coherence of an input states(except $c_{1}=c_{2}=0$) no mater what the memory coefficient of channel $ \mu$ is. This case is different from that of other noisy channels we consider following.

\subsection{Phase-damping channel with memory}
Following, we go on to study the quantum coherence dynamics of two qubits over phase-damping channel with memory. The Kraus operators for a single qubit are defined in terms of the Pauli operators $\sigma_{0}=I$ and $\sigma_{3}$.

Similarly, if consider two consecutive uses quantum dephasing channel, the Kraus operators of channel with memoryless are given~\cite{20c}
\begin{equation}
E_{i,j}=\sqrt{P_{i}P_{j}}\sigma_{i}\otimes \sigma_{j}
\end{equation}
and the Kraus operators of this channel with memory are given
\begin{equation}
E_{k,k}=\sqrt{P_{k}}\sigma_{k}\otimes \sigma_{k}
\end{equation}
where $i,j,k=0,3$, and $P_{0}=1-p$, $P_{3}=p$, where $p=1-\exp(-\gamma t)$.
According to Eq.~\eqref{Eq3},  the corresponding time-dependent
density matrix of the phase-damping channel with memory is
\begin{equation}\label{Eq9}
\varepsilon(\rho)=\frac{1}{4}(I\otimes I+\Sigma_{i=1}^{3}c_{i}^{'}\sigma_{i}\otimes \sigma_{i})
\end{equation}
where

$c_{1}^{'}=c_{1}[1 + (p-2) p (1 - \mu)]$,

$c_{2}^{'}=c_{2}[1 + (p-2) p (1 - \mu)]$,

$c_{3}^{'}=c_{3}$.

Using Eqs.~\eqref{Eq9}, Eq.~\eqref{Eq1} and Eq.~\eqref{Eq2},  we derive the expression of the $l_{1}$ norm
and relative entropy of coherence for two qubits through the phase-damping channel with memory
\begin{equation}\label{Eq10}
C_{l_1}(\varepsilon(\rho))=\frac{1}{2}(|c_{1}-c_{2}|+|c_{1}+c_{2}|)[1-(2-p)p(1-\mu)],
\end{equation}
and
\begin{eqnarray}\label{Eq11}
C_{R}(\varepsilon(\rho))&=&\frac{1}{4}\{2(c_{3}-1)\log_{2}\frac{1-c_{3}}{4}-2(1+c_{3})\log_{2}\frac{1+c_{3}}{4}\nonumber\\
&+&(1-c_{3}+c_{1}^{'}+c_{2}^{'})\log_{2}\frac{1}{4}(1-c_{3}+c_{1}^{'}+c_{2}^{'})\nonumber\\
&+&(1+c_{3}+c_{1}^{'}-c_{2}^{'})\log_{2}\frac{1}{4}(1+c_{3}+c_{1}^{'}-c_{2}^{'})\nonumber\\
&+&(1+c_{3}-c_{1}^{'}+c_{2}^{'})\log_{2}\frac{1}{4}(1+c_{3}-c_{1}^{'}+c_{2}^{'})\nonumber\\
&+&(1-c_{3}-c_{1}^{'}-c_{2}^{'})\log_{2}\frac{1}{4}(1-c_{3}-c_{1}^{'}-c_{2}^{'})\}
\end{eqnarray}
Fig.2 shows the evolutions of $l_{1}$ norm and relative entropy with respect to $p$ for different parameters $\mu$.
The results show quantum coherence can be protected effectively. As the memory coefficient of channel $\mu$ increases, quantum coherence can be protected more effectively. In the case of $\mu=1$, quantum coherence is unaffected by the phase damping channels. One can easy prove it from Eqs.~\eqref{Eq10} and ~\eqref{Eq11} and following the condition given in Ref.~\cite{19}
\begin{equation}\label{Eq15}
\partial_{p}C_{l_1}(\varepsilon(\rho))=\partial_{p}C_{R}(\varepsilon(\rho))=0
\end{equation}
which is unaffected by the phase noise for any $p\in [0,1]$ (namely for any $t$), only when either $\mu=1$ or $c_{1}=c_{2}=0$(the input state is incoherence). Hence, we prove coherence can be unaffected by the phase damping channels is that $\mu=1$. On the other hand, in the limited $p\rightarrow 1$ ($t\rightarrow \infty$), Eqs.~\eqref{Eq10} and ~\eqref{Eq11} reduce to
\begin{eqnarray}\label{Eq8a}
C_{l_1}(\varepsilon(\rho))&=&\frac{1}{2}\mu(|c_{1}+c_{2}|+|c_{1}-c_{2}|)
\end{eqnarray}
and
\begin{eqnarray}\label{Eq8b}
C_{R}(\varepsilon(\rho))&=&\frac{1}{4}\{2(c_{3}-1)\log_{2}(\frac{1-c_{3}}{4})-2(c_{3}+1)\log_{2}(\frac{1+c_{3}}{4})\nonumber\\
&+&[\mu(c_{1}-c_{2})+1+c_{3}]\log_{2}\frac{1}{4}[\mu(c_{1}-c_{2})+1+c_{3}]\nonumber\\
&+&[-\mu(c_{1}-c_{2})+1+c_{3}]\log_{2}\frac{1}{4}[-\mu(c_{1}-c_{2})+1+c_{3}]\nonumber\\
&+&[-\mu(c_{1}-c_{2})+1-c_{3}]\log_{2}\frac{1}{4}[-\mu(c_{1}-c_{2})+1-c_{3}]\nonumber\\
&+&[\mu(c_{1}-c_{2})+1-c_{3}]\log_{2}\frac{1}{4}[\mu(c_{1}-c_{2})+1-c_{3}]
\end{eqnarray}
which indicate long-live coherence is related to the initial input states and the memory coefficient of channel $ \mu$.

\subsection{Depolarizing channel with memory}
Finally, we will study the quantum coherence dynamics of two qubits over depolarizing channel, it describes the process in which the density matrix is dynamically replaced by the state $I/2$. $I$ denoting identity matrix of a qubit. The Kraus operators for a single qubit are given by~\cite{20d}
\begin{equation}
A_{i}=\sqrt{P_{i}}\sigma_{i}
\end{equation}
where $i,j,k=0,1,2,3$, and $P_{0}=1-p$, $P_{1}=P_{2}=P_{3}=p/3$, where $p=1-\exp(-\gamma t)$.

By definition, the Kraus operators of depolarizing channel with memoryless are described as
\begin{equation}
E_{i,j}=\sqrt{P_{i}P_{j}}\sigma_{i}\otimes \sigma_{j}
\end{equation}
and the case of two consecutive uses of a channel with partial memory,
the task of constructing Kraus operators $E_{k,k}$ is given
\begin{equation}
E_{k,k}=\sqrt{P_{k}}\sigma_{k}\otimes \sigma_{k}.
\end{equation}
According to Eq.~\eqref{Eq3},  the corresponding time-dependent
density matrix of the depolarizing channel with memory is
\begin{equation}\label{Eq12}
\varepsilon(\rho)=\frac{1}{4}(I\otimes I+\Sigma_{i=1}^{3}c_{i}^{'}\sigma_{i}\otimes \sigma_{i})
\end{equation}
where

$c_{1}^{'}=\frac{1}{9}c_{1}[9 + 8 p (2 p-3) (1 - \mu)]$,

$c_{2}^{'}=\frac{1}{9}c_{2}[9 + 8 p (2 p-3) (1 - \mu)]$,

$c_{3}^{'}=\frac{1}{9}c_{3}[9 + 8 p (2 p-3) (1 - \mu)]$.

Similarly, taking Eq.~\eqref{Eq12} into Eq.~\eqref{Eq1} and  Eq.~\eqref{Eq2} respectively, the $l_{1}$ norm
and relative entropy of coherence for two qubits through the depolarizing channel with memory are
\begin{equation}\label{Eq13}
C_{l_1}(\varepsilon(\rho))=\frac{1}{18}(|c_{1}-c_{2}|+|c_{1}+c_{2}|)[9\mu+(3-4p)^2(1-\mu)],
\end{equation}
and
\begin{eqnarray}\label{Eq14}
C_{R}(\varepsilon(\rho))&=&\frac{1}{4}\{2(-1-c_{3}^{'})\log_{2}\frac{1}{4}(1+c_{3}^{'})\nonumber\\
&+&2(-1+c_{3}^{'})\log_{2}\frac{1}{4}(1-c_{3}^{'})\nonumber\\
&+&(1-c_{3}^{'}+c_{1}^{'}+c_{2}^{'})\log_{2}\frac{1}{4}(1-c_{3}^{'}+c_{1}^{'}+c_{2}^{'})\nonumber\\
&+&(1+c_{3}^{'}+c_{1}^{'}-c_{2}^{'})\log_{2}\frac{1}{4}(1+c_{3}^{'}+c_{1}^{'}-c_{2}^{'})\nonumber\\
&+&(1+c_{3}^{'}-c_{1}^{'}+c_{2}^{'})\log_{2}\frac{1}{4}(1+c_{3}^{'}-c_{1}^{'}+c_{2}^{'})\nonumber\\
&+&(1-c_{3}^{'}-c_{1}^{'}-c_{2}^{'})\log_{2}\frac{1}{4}(1-c_{3}^{'}-c_{1}^{'}-c_{2}^{'})\}
\end{eqnarray}

The evolutions of $l_{1}$ norm and relative entropy of two qubits over depolarizing channel with respect to $p$ for different parameters $\mu$ are depicted in Fig. 3, from which we can see coherence decreases firstly and then increases. This means the depolarizing channel with memory destories coherence firstly and then creates coherence.  Besides, coherence can be protected effectively when the memory coefficient of channel $\mu$ increases. In the limited $p\rightarrow 1$, quantum coherence is still long-live determined from Eqs.~\eqref{Eq13} and ~\eqref{Eq14}, and depends on input state parameters $c_{i}$ and $\mu$. Similar to previous analysis, only when either $\mu=1$ or $c_{1}=c_{2}=0$(the input state is incoherence),
quantum coherence is unaffected by the depolarizing damping channels.

\section{Cohering power and decohering power of correlated channels}\label{sec:P}
To better understand the quantum channel to create or destroy coherence of input quantum states,
we adopt to the concepts proposed by Mani $et.$ $al.$~\cite{20} who introduced the cohering and de-cohering power of any quantum channels. The cohering power of a channel is the maximal amount of coherence that it creates when acting on a completely incoherent state.
\begin{equation}\label{CP1}
C^K_{C}(\varepsilon)= \max\limits_{\rho\in\mathcal{I}}\{C^K(\varepsilon(\rho))-C^K(\rho)\},
\end{equation}
$C^K_{C}$ denotes the cohering power of the coherence measure $C$, we here only consider the coherence measure $C_{l_1}$ and $C_{R}$. $K$ is the reference basis, for simplicity, we take Bell basis $K=\{|00\rangle, |01\rangle, |10\rangle, |11\rangle\}$ as the fixed basis on two-qubit system,

Based on Ref.~\cite{20}, Eq.~\eqref{CP1} can reduce
\begin{equation}\label{CP2}
C^K_{C}(\varepsilon)=\max\limits_{i}C^K(\varepsilon(|k_{i}\rangle\langle k_{i}|)).
\end{equation}

The decohering power of the channel $\varepsilon$ is the maximum amount by which it reduces the coherence of a maximally coherent state defining as~\cite{20}]
\begin{equation}\label{CP3}
D^K_{C}(\varepsilon)=\max\limits_{\rho\in M}\{C^K(\rho)-C^K(\varepsilon(\rho))\}.
\end{equation}
$M$ is a set of maximally coherent states. According to Ref.~\cite{21}, all maximally coherent states are pure ones, for instance, $M=\frac{1}{\sqrt{2}}(e^{i\alpha}|0\rangle+e^{i\beta}|1\rangle)\otimes \frac{1}{\sqrt{2}}(e^{i\theta}|0\rangle+e^{i\phi}|1\rangle)$, where $\alpha,\beta,\theta,\phi \in [0,2\pi)$.

\begin{figure}[htpb]
  \begin{center}
   \includegraphics[width=7.5cm]{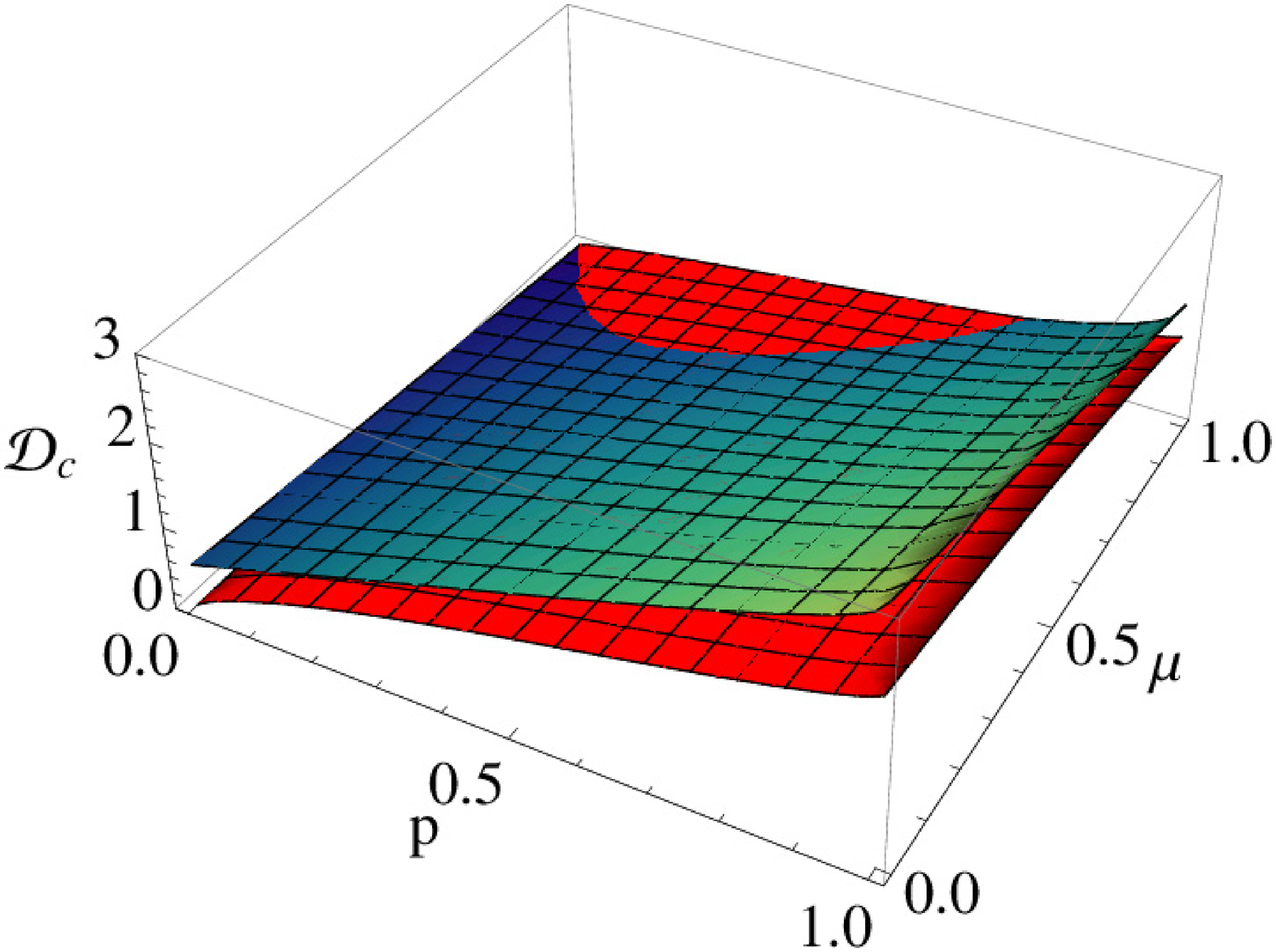}
   \includegraphics[width=7.5cm]{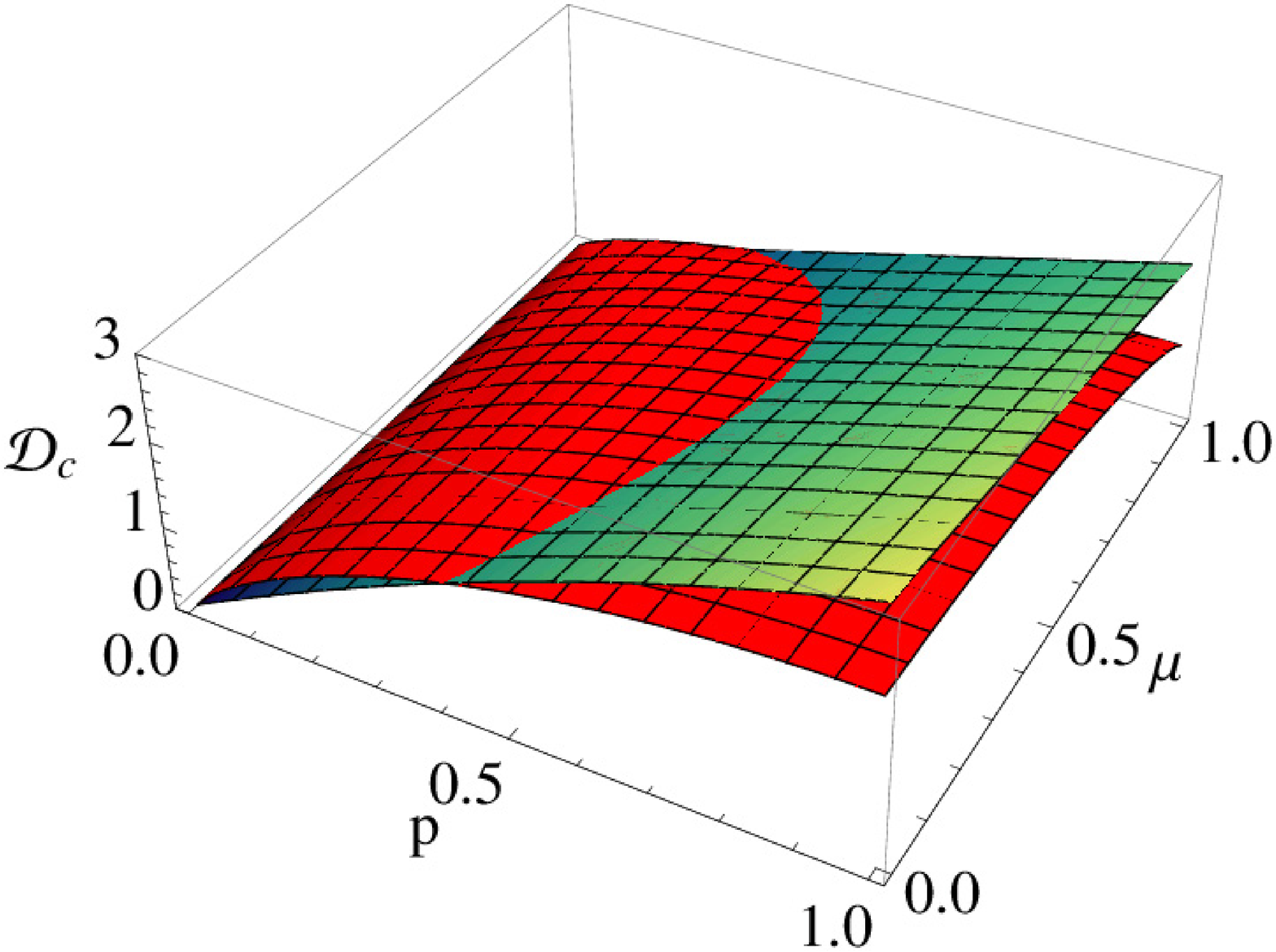}
   \includegraphics[width=7.5cm]{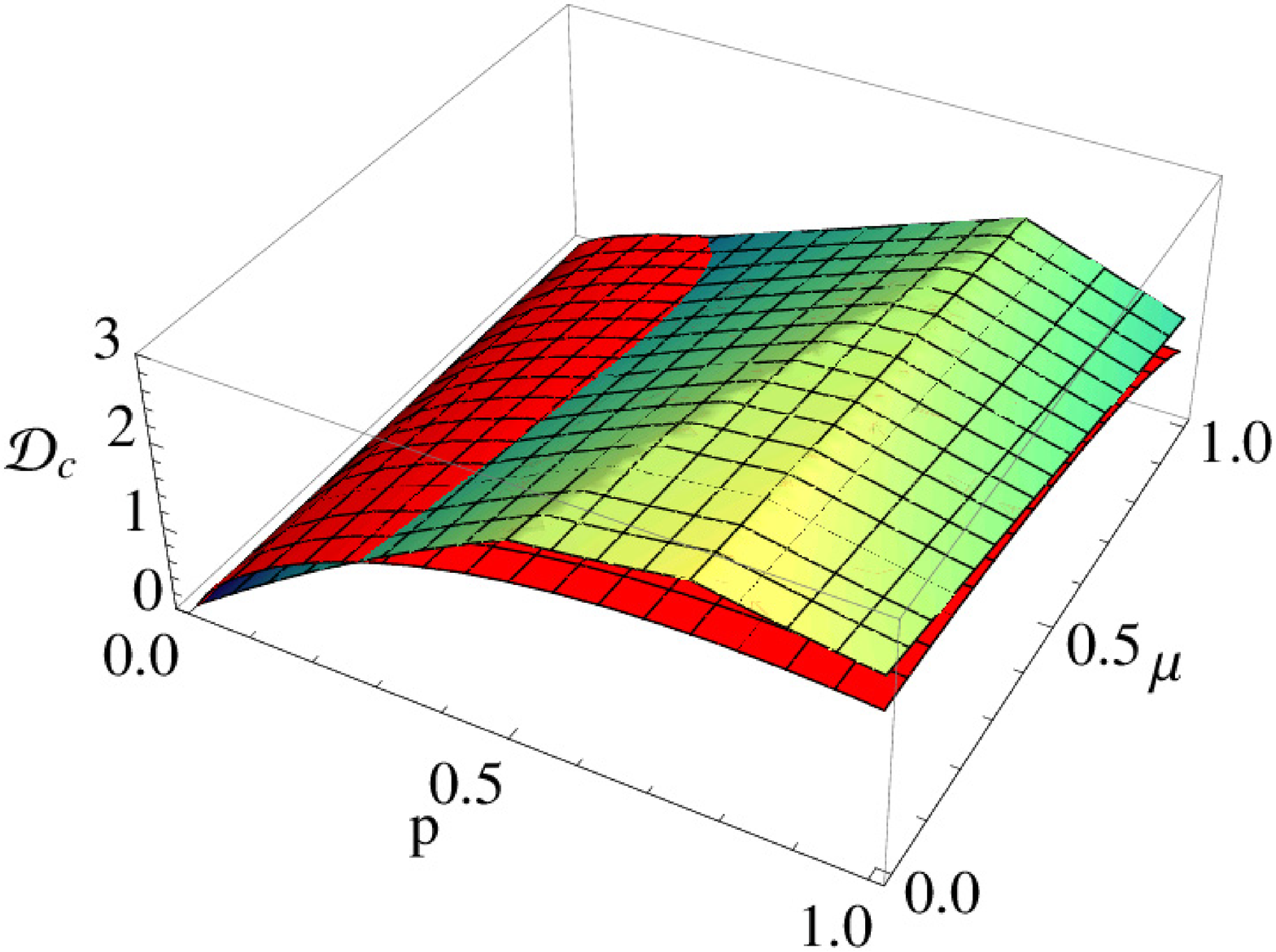}
 \caption{\label{Fig4}(Color online) Decohering power of different quantum channels with memory respect to the $l_{1}$ norm (Upper color surface) and the relative entropy (Lower red surface). From top to bottom (a)Amplitude damping channels, (b)phase damping channels, (c)depolarizing damping channels.}
\end{center}
\end{figure}
Following, we resort to Eq.~\eqref{CP2} and~\eqref{CP3} respectively to study the cohering and decohering power of quantum channels with memory on the fixed basis $K=\{|00\rangle, |01\rangle, |10\rangle, |11\rangle\}$. Firstly, we consider the cohering power of the amplitude damping channel, phase damping channels, and depolarizing damping channels with memory. According to Eq~\eqref{CP2},
\begin{equation}\label{CP4}
C^K_{C}(\varepsilon)= 0,
\end{equation}
where $C$ donates either the $l_{1}$ norm or the relative entropy to measure coherence, and $K$ corresponds to the fixed basis. Eq.~\eqref{CP4} means that those noisy channels don't have any cohering power with respect ro the fixed basis $K=\{|00\rangle, |01\rangle, |10\rangle, |11\rangle\}$.

According to Eq~\eqref{CP3}, the decohering power of the amplitude damping channel on the fixed basis$K$
\begin{equation}\label{Eq:DP5}
D_{C_{l}}(\varepsilon)=3-\frac{1}{2}\{2+3\sqrt{1-p}-(2+\sqrt{1-p})p(1-\mu)+\mu\},
\end{equation}
Since the analytical expression of the decohering power of the amplitude damping channel with respect to the relative entropy $D_{C_{R}}$ on the fixed basis$K$ is complexity, we here make numerical results.
Similarly, the decohering power of the dephasing damping channel
\begin{equation}\label{Eq:DP6}
D_{C_{l}}(\varepsilon)=3-[1-p^2(1-\mu)-2p\mu]
\end{equation}
and
\begin{eqnarray}\label{Eq:DP7}
D_{C_{R}}(\varepsilon)&=&\frac{1}{4}\{(p-2)[(2-p+p\mu)\log_{2}\frac{1}{4}(2-p)(2-p+p\mu)\nonumber\\
&+&2p(1-\mu)\log_{2}\frac{1}{4}(2-p)p(1-\mu)]\nonumber\\
&-&(p^2(1-\mu)+2\mu)\log_{2}\frac{1}{4}p(p+2\mu-p\mu)\},
\end{eqnarray}
the decohering power of the depolarizing damping channel
\begin{eqnarray}\label{Eq:DP8}
D_{C_{l}}(\varepsilon)&=&3-\frac{1}{9}\{9-4(3-p)p(1-\mu)\nonumber\\
&+&3[|3-4p|+|3-4p(2-p-\mu+p\mu)|]\}
\end{eqnarray}
and
\begin{eqnarray}\label{Eq:DP9}
D_{C_{R}}(\varepsilon)&=&\frac{1}{9}\{2p(p-p\mu-3)\log_{2}\frac{1}{9}p[3-p(1-\mu)]\nonumber\\
&-&(3-2p)(3-2p+2p\mu)\log_{2}\frac{1}{9}(3-2p)(3-2p+2p\mu)\nonumber\\
&+&2p(1-\mu)(3-2p)[\log_{2}\frac{9}{2}-\log_{2}p(3-2p)(1-\mu)]\nonumber\\
&-&2p(1-\mu)[p\log_{2}\frac{2}{9}p^2(1-\mu)]\}
\end{eqnarray}
Fig.4 shows decohering power of different quantum channels with memory respect to the $l_{1}$ norm and the relative entropy on the fixed basis $K$. It can be seen clearly they all increased monotonically with $P$ from 0 to the maximal value, and decreased
monotonically with $\mu$.
\section{Conclusion}  
In this paper, we have investigated the dynamics of two-qubit quantum coherence in terms of the $l_{1}$ norm and the relative entropy when two qubits go through quantum channels where successive uses of the channel are correlated. Some analytical or numerical results are presented. The results show that, the quantum coherence of two-qubit system can be well protected. The stronger the memory coefficient of channel is, the much more slowly the quantum coherence decay.  Particularly, as the phase damping and depolarizing channels with perfect memory, quantum coherence is unaffected by noisy channels. Finally, we discuss the cohering and de-cohering powers of correlated channels with respect to the $l_{1}$ norm and relative entropy on the fixed basis. These results may be useful to the study of coherence.

\acknowledgments
This research is supported by the Start-up Funds for Talent Introduction and Scientific Research of Changsha University 2015 (SF1504) and Scientific Research Project of Hunan Province Department of Education (16C0134) and the Project of Science and Technology Plan of Changsha (ZD1601071)

\label{app:eff-trans}

\end{document}